\documentclass[twocolumn, trackchanges]{aastex631}
\usepackage{amsmath}
\usepackage[normalem]{ulem}
\submitjournal{APJL}

\begin{document}

\title{Detection of Oscillations in a Type-I X-Ray Burst of 4U~0614+091 with SVOM/ECLAIRs}

\author[0000-0002-2501-5589]{Sébastien Le Stum}
\correspondingauthor{Sébastien Le Stum}
\email{lestum@apc.in2p3.fr}
\affiliation{Université Paris Cité, CNRS, Astroparticule et Cosmologie, F-75013 Paris, France}

\author{Floriane Cangemi}
\author{Alexis Coleiro}
\affiliation{Université Paris Cité, CNRS, Astroparticule et Cosmologie, F-75013 Paris, France}

\author{Sébastien Guillot}
\affiliation{IRAP, Université de Toulouse, CNRS, CNES, 9 avenue du Colonel Roche, BP 44346, F-31028 Toulouse Cedex 4, France}

\author{Jérôme Chenevez}
\affiliation{DTU Space, Technical University of Denmark, Elektrovej 327-328, DK-2800 Kongens Lyngby, Denmark}

\collaboration{30}{ }

\author{Philippe Bacon}
\affiliation{Université Paris Cité, CNRS, Astroparticule et Cosmologie, F-75013 Paris, France}

\author{Nicolas Bellemont}
\affiliation{Université Paris Cité, CNRS, Astroparticule et Cosmologie, F-75013 Paris, France}

\author{Laurent Bouchet}
\affiliation{IRAP, Université de Toulouse, CNRS, CNES, 9 avenue du Colonel Roche, BP 44346, F-31028 Toulouse Cedex 4, France}

\author{Tristan Bouchet}
\affiliation{Université Paris Cité, Université Paris-Saclay, CEA, CNRS, AIM, 91191 Gif-sur-Yvette, France}

\author{Cécile Cavet}
\affiliation{Université Paris Cité, CNRS, Astroparticule et Cosmologie, F-75013 Paris, France}

\author{Bertrand Cordier}
\affiliation{Université Paris-Saclay, Université Paris Cité, CEA, CNRS, AIM, 91191 Gif-sur-Yvette, France}

\author{Antoine Foisseau}
\affiliation{Université Paris Cité, CNRS, Astroparticule et Cosmologie, F-75013 Paris, France}

\author{Olivier Godet}
\affiliation{IRAP, Université de Toulouse, CNRS, CNES, 9 avenue du Colonel Roche, BP 44346, F-31028 Toulouse Cedex 4, France}

\author{Andrea Goldwurm}
\affiliation{Université Paris Cité, CNRS, CEA, Astroparticule et Cosmologie, F-75013 Paris, France}
\affiliation{CEA Paris-Saclay, Irfu/Département d’Astrophysique, 91191 Gif-sur-Yvette, France}

\author{Xuhui Han}
\affiliation{National Astronomical Observatories, Chinese Academy of Sciences, Beijing 100101, China}

\author{Cyril Lachaud}
\affiliation{Université Paris Cité, CNRS, Astroparticule et Cosmologie, F-75013 Paris, France}

\author{Zhaosheng Li}
\affiliation{Key Laboratory of Stars and Interstellar Medium, Xiangtan University, Xiangtan 411105, Hunan, China}

\author{Huali Li}
\affiliation{National Astronomical Observatories, Chinese Academy of Sciences, Beijing 100101, China}

\author{Yulei Qiu}
\affiliation{National Astronomical Observatories, Chinese Academy of Sciences, Beijing 100101, China}

\author{Jérôme Rodriguez}
\affiliation{Université Paris-Saclay, Université Paris Cité, CEA, CNRS, AIM, 91191 Gif-sur-Yvette, France}

\author{Wenjun Tan}
\affiliation{Institute of High Energy Physics, Chinese Academy of Sciences, 19B Yuquan Road, Beijing 100049, China}

\author{Lian Tao}
\affiliation{Institute of High Energy Physics, Chinese Academy of Sciences, 19B Yuquan Road, Beijing 100049, China}

\author{Lauryne Verwaerde}
\affiliation{Université Paris Cité, CNRS, Astroparticule et Cosmologie, F-75013 Paris, France}

\author{Chenwei Wang}
\affiliation{Institute of High Energy Physics, Chinese Academy of Sciences, 19B Yuquan Road, Beijing 100049, China}

\author{Jing Wang}
\affiliation{National Astronomical Observatories, Chinese Academy of Sciences, Beijing 100101, China}

\author{Jianyan Wei}
\affiliation{National Astronomical Observatories, Chinese Academy of Sciences, Beijing 100101, China}

\author{Chao Wu}
\affiliation{National Astronomical Observatories, Chinese Academy of Sciences, Beijing 100101, China}

\author{Wenjin Xie}
\affiliation{National Astronomical Observatories, Chinese Academy of Sciences, Beijing 100101, China}

\author{Liping Xin}
\affiliation{National Astronomical Observatories, Chinese Academy of Sciences, Beijing 100101, China}

\author{Shaolin Xiong}
\affiliation{Institute of High Energy Physics, Chinese Academy of Sciences, 19B Yuquan Road, Beijing 100049, China}

\author{Shuangnan Zhang}
\affiliation{Institute of High Energy Physics, Chinese Academy of Sciences, 19B Yuquan Road, Beijing 100049, China}

\author{Shijie Zheng}
\affiliation{Institute of High Energy Physics, Chinese Academy of Sciences, 19B Yuquan Road, Beijing 100049, China}



\begin{abstract}

On 2025 January 10, a thermonuclear (Type I) X-ray burst from the neutron star low-mass X-ray binary \textit{4U~0614+091} was detected with the ECLAIRs instrument on board the \textit{SVOM} mission. We present here a time-resolved spectroscopic analysis of the burst, along with the detection of burst oscillations within a 51-second interval during the decay phase. The oscillation frequency is measured to be $\nu = 413.674 \pm 0.002\,\mathrm{Hz}$, consistent with previous reports. However, we detect a significant downward frequency drift over the burst duration, characterized by $\dot{\nu} = (-4.7 \pm 0.3) \times 10^{-3}\,\mathrm{Hz\,s^{-1}}$. This frequency evolution is atypical compared to those observed in similar burst oscillation sources. We tentatively attribute the observed drift to a Doppler shift induced by orbital motion. Under this interpretation, the inferred orbital period must be shorter than 20 minutes, placing \textit{4U~0614+091} among the most compact known low-mass X-ray binaries.

\end{abstract}



\section{Introduction} 
\label{sec:intro}

Low-mass X-ray binaries (LMXBs) hosting neutron star accretors are known to exhibit intense, short-duration outbursts resulting from thermonuclear fusion of accreted material on the neutron star surface. These bursts typically present lightcurves characterized by a rapid rise in flux followed by a cooling tail lasting tens of seconds \citep{Galloway_2020_minbar}. In some cases, thermonuclear bursts exhibit coherent oscillations, referred to as burst oscillations \citep{1996ApJ...469L...9S}. In systems where a persistent pulsation is also observed, burst oscillation frequencies are found to be close to the spin frequency of the neutron star, indicating a strong connection between the two phenomena \citep[see][for a review]{2012ARA&A..50..609W}.

The LMXB 4U~0614+091 is an ultracompact system composed of a neutron star and a white dwarf companion. The distance to it has been estimated to be $3.3^{+1.3}_{-2.4}$~kpc based on the GAIA parallax of its optical counterpart \citep{2021MNRAS.502.5455A}; or $2.59\pm 0.03$~kpc under the hypothesis that a photospheric radius expansion burst detected in 2001 was emitting at the Eddington luminosity with a pure He composition \citep{Galloway_2020_minbar}.
In the following, an approximate distance of 3~kpc is chosen for the relevant computations.
Evidence of quasi-periodic oscillations \citep[QPO,][]{Ford_1997} and of an accretion-powered compact radio jet  \citep{Migliari_2010} have been observed. Constraints on the neutron star mass, derived from the maximum observed QPO frequency, suggest a mass of approximately $2\,M_\odot$ \citep{10.1111/j.1365-2966.2009.15430.x,10.1093/mnras/sty1404}.
Spectroscopic studies indicate that the accretion flow from the white dwarf is rich in carbon and oxygen but deficient in hydrogen and helium \citep{Nelemans_2004, Madej_2014}. However, the observed burst recurrence rate implies that a non-negligible fraction of helium must be present in the accreted material \citep{2010A&A...514A..65K}. 
Despite extensive study, the orbital period of 4U~0614+091 remains uncertain. Optical photometry has revealed a modulation with a period of approximately 50 minutes \citep{2008PASP..120..848S}, while \citet{2013MNRAS.429.2986M} report a periodicity of approximately 30~minutes in the 4650\,\r{A} oxygen line. Alternatively, \citet{2014A&A...572A..99B} argue for an orbital period~$\gtrsim$\,1\,hour, based on the assumption that the modulation arises from the outer rim of the accretion disk.

Burst oscillations at a frequency of 414.75~Hz were previously detected in 4U~0614+091 during an October 2006 burst observed with the Swift/BAT instrument \citep{bat_discovery}. Additionally, a possible oscillation signal near 413~Hz was reported from a burst detected by the GECAM instrument \citep{gecam_detection} in January 2021.

In this Letter, we present the detection of burst oscillations at a frequency of $413.674\pm 0.002$~Hz using the ECLAIRs X-ray telescope on board the Space-based multi-band astronomical Variable Object Monitor (SVOM) satellite. Section \ref{sec:data} provides a brief overview of the instrument and the analysis methods. The results of the time-resolved spectroscopy and oscillation search are presented in Section~\ref{sec:results}, followed by a discussion of their interpretation in Section~\ref{sec:discussion} and conclusions in Section~\ref{sec:conclusion}.

\section{Data set}
\label{sec:data}

SVOM is a Chinese-French space mission, launched on June 22$^{nd}$, 2024, designed to explore the high-energy transient sky \citep{wei2016}.  The satellite is equipped with four instruments: the Microchannel X-Ray Telescope (MXT), a soft X-ray (0.2 to 10 keV) instrument using micro-pore optics (lobster eyes, \citealt{gotz_2014}); the Gamma Ray Burst Monitor (GRM), a set of three detectors sensitive to photons in the 15\,keV to 5\,MeV range \citep{Dong2010}; the Visible Telescope \citep[VT,][]{wu_2012}, a 40\,cm optical telecope with two channels (blue and red); and ECLAIRs, a large field of view (FoV) of 89\textdegree~by 89\textdegree~coded-mask telescope \citep{Goldwurm_2022} operating in the 4 to 150 keV range with a 10\,$\mu$s timing resolution \citep{Godet_2014}. The satellite's primary objective is the detection and characterisation of Gamma-Ray Bursts. To this end, ECLAIRs operates an onboard triggering system that identifies fast transient events and determines their sky position. If the trigger corresponds to a source not listed in the onboard catalog, the system can autonomously request a rapid reorientation (slew) of the spacecraft to enable follow-up observations \citep{Dagoneau_2020, Dagoneau_2022}.

On 2025-01-10, the ECLAIRs onboard system detected a significant excess with a signal-to-noise ratio of 47 in an image integrated over a 20.48-second interval, starting at $T_0~=~\text{2025-01-10T15:58:03 UTC}$, in the 5–8 keV energy band. This detection triggered an alert with coordinates consistent with the known source 4U~0614+091, located 22.3\textdegree~from the center of the FoV. As the source was listed in the onboard catalog, no satellite slew was initiated in response to the trigger. At the time of the trigger, the Earth was outside the telescope’s FoV, and the spacecraft was approaching the South Atlantic Anomaly (SAA), resulting in an increasing background count rate due to charged particle flux.

\begin{figure}[t]
    \centering
    \includegraphics[width=\linewidth]{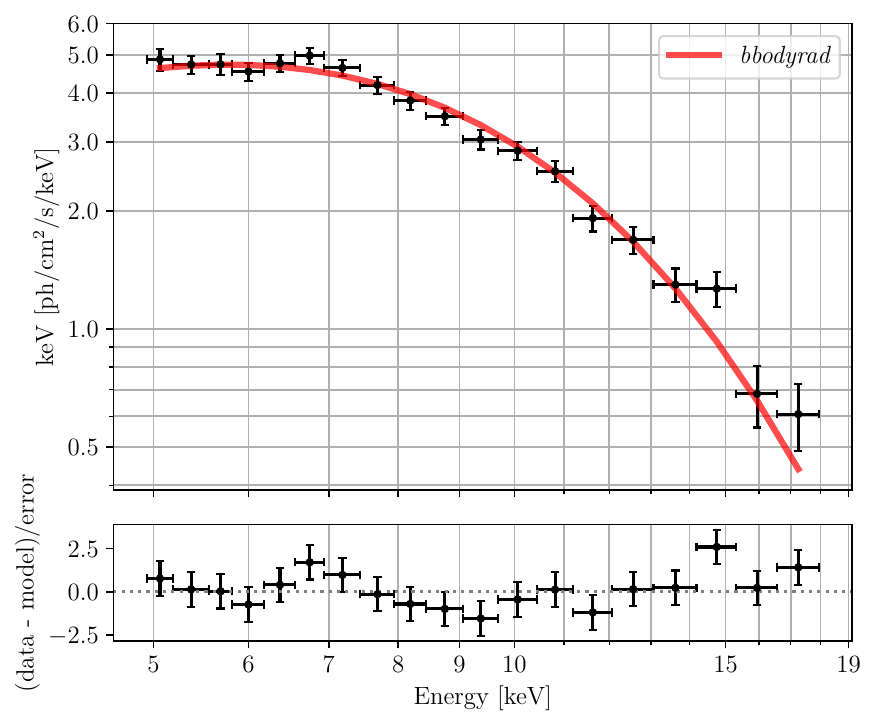}
    \caption{Time-integrated spectrum between $T=T_0$ and $T=T_0 +25$\,s, with in red a fitted \textit{bbodyrad} model with $kT = 2.04$~keV.}
    \label{fig:avg_spectrum}
\end{figure}

\subsection{Data analysis}

The reduction and analysis of ECLAIRs data for 4U~0614+091 were performed using the ECLAIRs Pipeline (ECPI), developed for the mission science ground segment (Goldwurm et al., in preparation), which enables both imaging and spectroscopic analysis from event data.
As a first step, good time intervals (GTIs) were selected based on satellite attitude stability and instrumental conditions.
Around the relevant period for this work, the GTIs were uninterrupted.
Calibrated detector-plane images were then generated in a set of energy bins between 4 and 30 keV.
Detector images in a few large energy bins were spatially deconvolved using the coded-mask pattern to produce sky images. A point-source extraction algorithm was subsequently applied to the sky images to fit the positions of all sources within the FoV. 
Using the fitted position ensures the accurate modeling of sources for the spectral extraction and event selection, avoiding effects of biases in the telescope orientation with respect to the satellite's attitude.
Sources count rates were derived in 30 smaller, logarithmically spaced, energy bins by modeling the contributions to each detector pixel from both 4U~0614+091 and the Crab Nebula, the only other bright source present in the FoV, and simultaneously fitting them to the binned detector images. A background component, modeled as a 2-D  polynomial, was also included in the fit.

\section{Results}
\label{sec:results}

\subsection{Spectroscopy}
\label{subsec:spectrum}

The source count rate spectrum, derived by the procedure described above, was fitted using \textsc{Xspec} version 12.15.0 \citep{1996ASPC..101...17A}, with a $\chi^2$ minimization method. The uncertainties on spectral parameters were computed at the 90\% confidence level. A time-integrated spectrum, covering the interval from $T_0$ to $T_0 + 25$~s, is shown in \autoref{fig:avg_spectrum}, where the best-fit \texttt{bbodyrad} model is overlaid on the data. 
Applying a 2\% systematic uncertainty, the fit yields a reduced chi-squared of $\chi^2_r~=~19.7/17~=~1.16$. The inclusion of an additional power-law component does not significantly improve the fit, meaning that we are unable to study the non-thermal contribution to the spectrum. 
Residuals in the spectrum show some features, but it is unclear whether these arise from still uncorrected instrumental systematics or from physical effects, such as absorption edges similar to those reported in GRS~1747–312 by \citet{Li_2018}. 
The best-fit parameters are a blackbody temperature of $kT = 2.04 \pm 0.04$ keV and an apparent emission radius of $R = 5.8 \pm 0.3$ km, assuming a source distance of 3 kpc. From this model, we derive a time-integrated bolometric burst energy of $E_{\mathrm{bol}} = (1.89 \pm 0.04) \times 10^{39}$~erg, corresponding to a fluence of $F_{\mathrm{bol}} = (1.75 \pm 0.04) \times 10^{-6}$~erg~cm$^{-2}$. These values are consistent with typical bursts from this source, as reported in Table 2 of \citet{2010A&A...514A..65K}. 

The source exhibits a significant accretion emission around the burst. Between 2025-01-10T13:46:13 (start of the observation) and 2025-01-10T15:57:52 ($T_0-11$\,s), with an exposure of 2.7~ks, its spectrum is well described ($\chi^2_r = 5.5/7 = 0.79$) by a power-law with a photon index of $2.65\pm0.20$ and a flux between 4 and 30~keV of $1.2^{+0.1}_{-0.4} \times 10^{-9}$~erg~cm$^{-2}$~s$^{-1}$. After the burst, between 2025-10-01T16:48:34 and 2025-10-01T21:38:25 (end of the observation), with an exposure of 9.6~ks, its photon index is $2.97\pm0.12$ and its 4-30 keV flux is $1.41^{+0.04}_{-0.19} \times 10^{-9}$~erg~cm$^{-2}$~s$^{-1}$ ($\chi^2_r = 7.4/6 = 1.23$).
On a longer timescale, between 2024 October 23 and 2025 March 3, the source was within the ECLAIRs FoV for a total of 117~ks of GTIs, excluding the observations previously mentioned. The time-integrated persistent emission spectrum is well described by a power-law with a photon index of $2.77\pm0.12$, yielding a reduced chi-squared of $\chi^2_r = 4.7/11 = 0.43$. The corresponding average persistent flux in the 4–30 keV range is $1.48^{+0.07}_{-0.16} \times 10^{-9}$~erg~cm$^{-2}$~s$^{-1}$. 
No additional Type I bursts were detected by ECLAIRs during this period.

\begin{figure}[t]
    \centering
    \includegraphics[width=\linewidth]{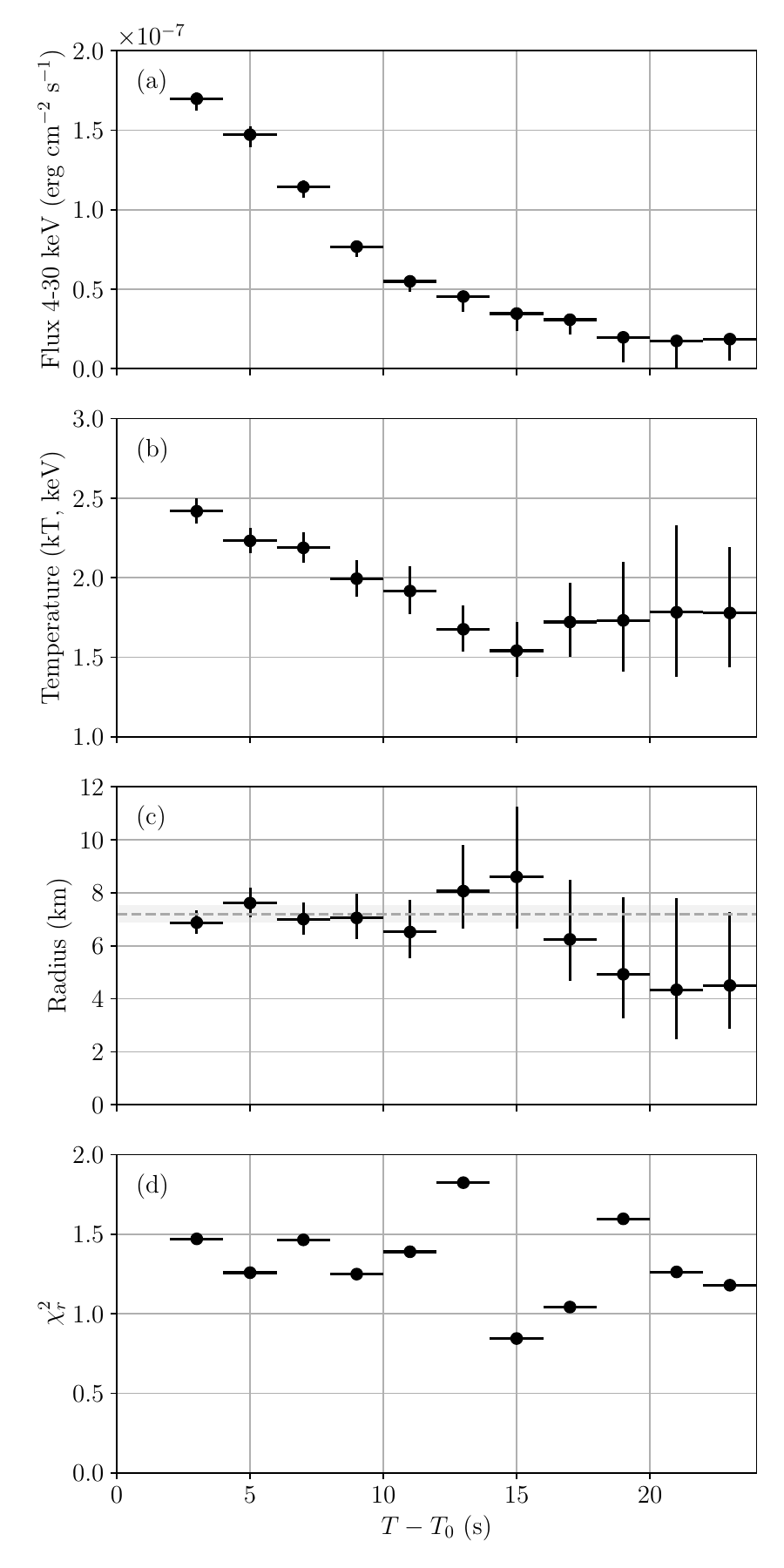}
    
    \caption{Evolution of time-resolved spectroscopy with a \textit{bbodyrad} model, with \textbf{(a)} the flux between 4 and 30 keV, \textbf{(b)} the temperature $kT$, \textbf{(c)} the emission radius, assuming a distance of 3 kpc, with a constant radius shown as a dashed line, and \textbf{(d)} the reduced chi square $\chi^2_r=\chi^2/17$ degrees of freedom.
    \label{fig:spectrum_evolution}}
\end{figure}

The results of a time-resolved spectral analysis during the burst are shown in \autoref{fig:spectrum_evolution}, which presents the evolution of the flux, blackbody temperature, and apparent emission radius, assuming a source distance of 3 kpc. The analysis was performed using 2-second time bins starting from $T_0$. 
The 4-30 keV flux peaks at $1.70~\pm~0.04~\times~10^{-7}$~erg~cm$^{-2}$~s$^{-1}$, which is two orders of magnitude higher than the persistent flux, and is equivalent to 7.1 times the Crab flux in this energy range. 
The associated peak bolometric luminosity is $(2.09 \pm0.07)\times 10^{38}$~erg~s$^{-1}$, still assuming a distance of 3 kpc. 
A clear cooling trend is observed between $T_0 + 3$~s and $T_0 + 15$~s, with the blackbody temperature decreasing from $kT = 2.43 \pm 0.08$ keV to $kT = 1.60 \pm 0.09$ keV.
Beyond this point, the temperature appears to reach a plateau; however, due to the lower flux levels, it becomes difficult to identify any significant trend. 
No evidence of photospheric radius expansion was found, as the radius remains constant within error bars.
After $T_0 + 25$~s, the source is too faint to allow for a reliable spectral modeling. 

\begin{figure}[ht]
    \centering
    \includegraphics[width=\linewidth]{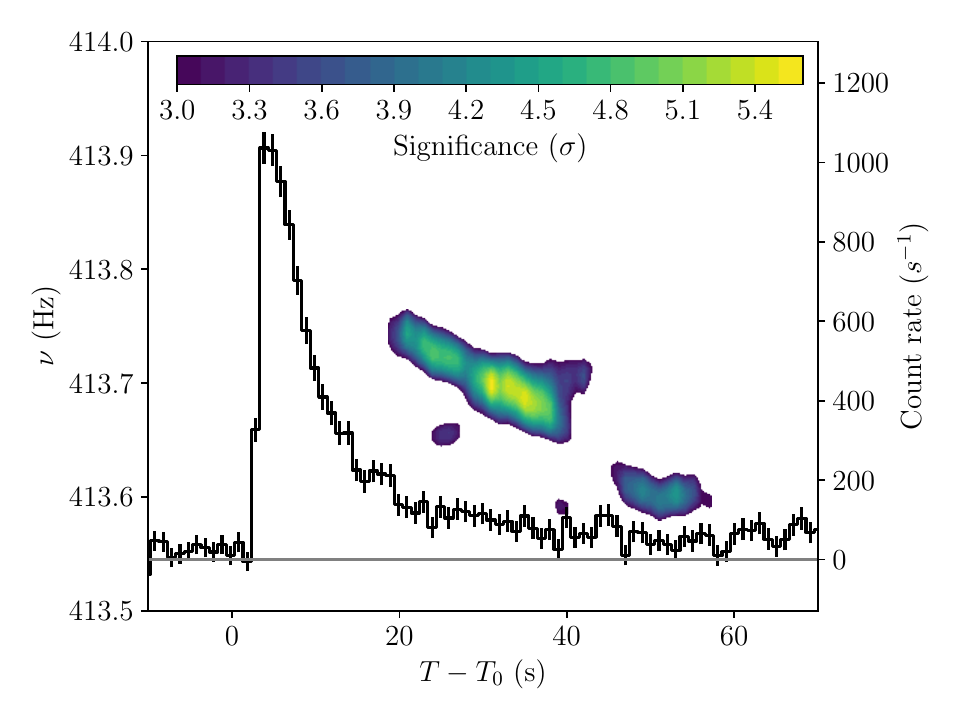}
    \caption{Significance contours of the oscillations as a function of time and frequency (left-hand axis), computed using a ${Z_1^2}$ test in a 20-second sliding window with 1-second steps. Significance levels are given as a color gradient. The burst light curve is shown as the equivalent on-axis count rate (right-hand axis) in the 4–40~keV energy band, with the persistent emission subtracted, binned in 1-second intervals.}
    \label{fig:contours_20s}
\end{figure}

\subsection{Burst Oscillations}
\label{subsec:oscillations}

Contrary to the spectral analysis, where the contribution of each source is obtained from the number of counts in each pixel, an oscillation search requires studying the arrival times of photons at an individual level. Thus, a selection of photons with respect to their position on the detector plane is performed to maximize the signal-to-noise ratio. 
Two primary sources of background noise in the timing analysis are (i) the elevated particle background due to the spacecraft’s proximity to the SAA at the time of the burst, and (ii) the presence of the Crab Nebula within the ECLAIRs FoV. The coded-mask pattern casts a shadow on the detector plane, such that each pixel has an illumination fraction, defined as 1 for full illumination by a source and 0 for full occultation by the mask.
To obtain a clean sample of photons for the oscillation search, we selected events from detector pixels with an illumination fraction greater than 0.9 for 4U~0614+091 and less than 0.1 for the Crab. 
Additionally, events selected within the 4–40 keV energy range. The lower bound corresponds to the instrument's detection threshold, while the upper bound maximizes the signal-to-noise ratio, as the thermal spectrum of the burst is not expected to contribute significantly at higher energies.

\begin{figure*}[htbp]
    \centering
    \includegraphics[width = 0.49\linewidth]{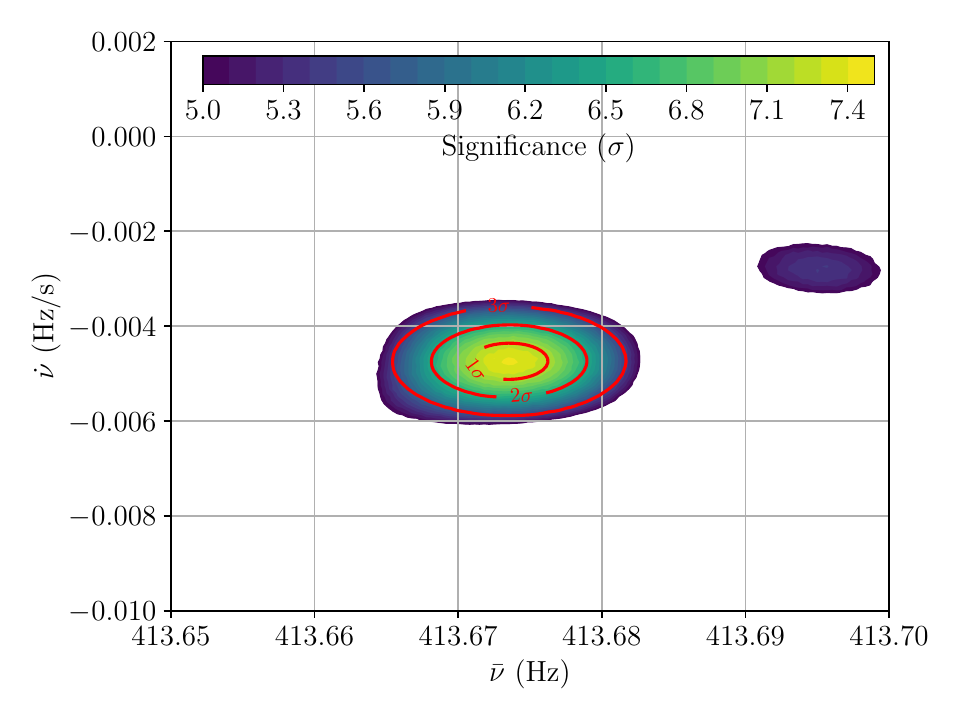}
    \includegraphics[width = 0.49\linewidth]{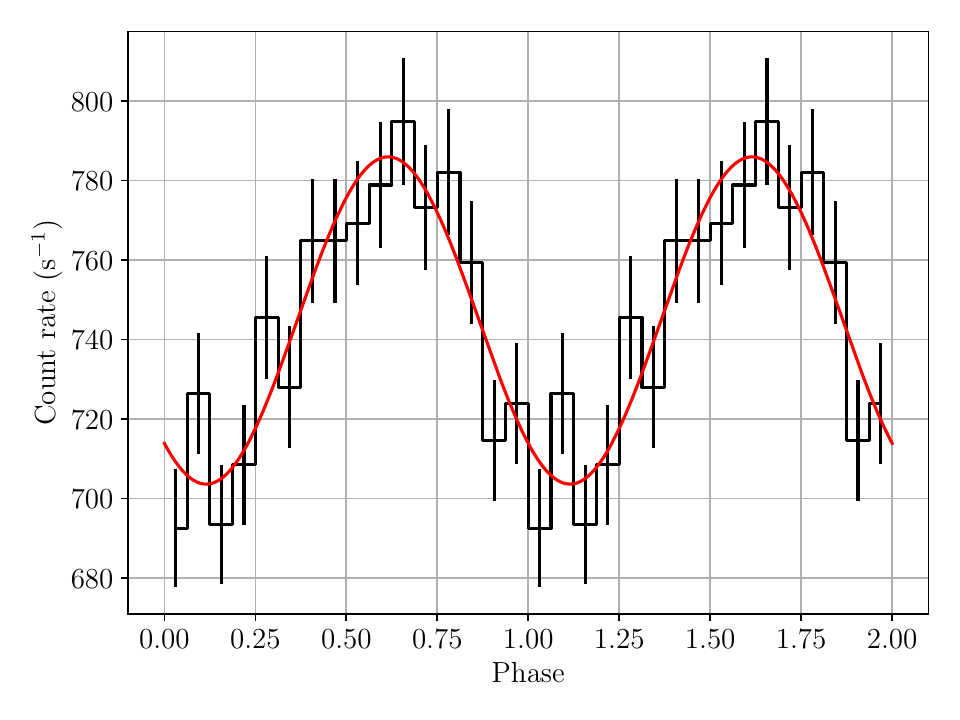}
    \caption{\textbf{Left:} Significance contours in the $\bar\nu$–$\dot\nu$ plane for the time window yielding the most significant signal [$T_0$+11~s,~$T_0$+62~s]. Significance levels are given as a color gradient. Overlaid in red are the contours of a two-dimensional Gaussian fitted around the peak. The best-fit parameters, $\bar\nu = 413.674$~Hz and $\dot\nu = -4.7 \times 10^{-3}$~Hz/s, are used to fold the light curve in this interval.
\textbf{Right:} Folded burst profile using the fitted parameters over the selected time window, in black. The best-fit sinusoidal model to the profile is shown in red}.
    \label{fig:nu_nudot}
\end{figure*}

Photon arrival times were corrected to the Solar System barycenter using the \textit{light\_travel\_time} routine from the \texttt{astropy} library version 6.1.7 \citep{2022ApJ...935..167A}. The position of the satellite is recorded as a housekeeping data stream sampled every 1 second. The time correction difference between each sample is at most $3.4 \times 10^{-5}$ s. This time resolution is sufficient to search for oscillations with periods much longer than this sampling interval, without requiring interpolation of the spacecraft’s position. Additionally, the radial velocity of the satellite relative to the source remains nearly constant over the short burst interval (tens of seconds). As a result, any residual error in the barycentric correction would introduce only a small Doppler shift in the oscillation frequency, without significantly affecting the coherence or detectability of the signal.

The search for burst oscillations is performed using the ${Z_n^2}$ statistics \citep{Buccheri:1983zz} implemented in the \textit{stingray} package, version 2.2.6 \citep{bachettiStingrayFastModern2024}.  
Under the null (background-only) hypothesis, the ${Z_n^2}$ test follows a $\chi^2$ distribution with $2n$ degrees of freedom, where $n$ is the number of harmonics considered in the test. From this assumption, the probability of obtaining a given ${Z_n^2}$ value under the null hypothesis can be computed. The final significance (p-value) is then obtained by accounting for the number of independent trials performed in the frequency search.

Following the previous findings from Swift/BAT \citep{bat_discovery} and GECAM \citep{gecam_detection} of significant burst oscillations at 414.7~Hz and 413.63~Hz, respectively, we searched for oscillations between 413 and 415~Hz using a resolution of 0.01~Hz.
Assuming that the oscillating signal may only be present during a portion of the burst, we searched for the time window and the frequency that maximize a $Z_1^2$ test between $T_0-5$\,s and $T_0+70$\,s in this frequency range.
Time windows are defined by varying their start and end times in 1-second increments, yielding a total of 2925 distinct intervals. 
We find the most significant signal in the time window [$T_0 + 27$\,s, $T_0 + 43$\,s] with a frequency of $\nu=$\,413.69\,Hz. The corresponding pre-trial p-value is $4.86 \times 10^{-9}$ (5.85$\,\sigma$). After accounting for the number of trials across both frequency and time dimensions, the post-trial p-value is $2.41 \times 10^{-3}$ (3.03$\,\sigma$ significance).
We note that this post-trial significance is conservative, as it assumes all trials are statistically independent.

\autoref{fig:contours_20s} shows the significance contours of the detected oscillation, computed using a 20-second sliding time window and overlaid on the burst light curve with the persistent emission subtracted.  The persistent level is estimated by extrapolating the pre-burst count rate, thus including contributions from both the background and any steady emission from 4U~0614+091. This subtraction is performed for visualization purposes and does not affect the timing analysis itself.
 \autoref{fig:contours_20s} shows a possible decrease in the oscillation frequency over time, which we examine and discuss in Section \ref{sec:drift}.

\subsection{Oscillation Frequency Drift}\label{sec:drift}

To investigate the possibility of a frequency drift during the burst, we introduced the frequency derivative $\dot{\nu}$ as an additional parameter in the burst oscillation search.  The phase of each photon with arrival time $t_j$ is then defined as $\phi_j = 2\pi(\nu t_j + \dot\nu t^2_j/2)$, which is then used to compute the ${Z_1^2}$ statistic.\footnote{The standard search described previously corresponds to the special case where $\dot{\nu} = 0$.} A validation of this analysis procedure using the Crab pulsar is described in \autoref{app:crab}.
We scanned $\dot{\nu}$ in the range from $-0.01$\,Hz/s to $+0.01$\,Hz/s in steps of 0.001\,Hz/s.
We find the most significant signal in the time window [$T_0$ + 11\,s, $T_0$ + 62\,s] with a frequency at $T=T_0$ + 11\,s of $\nu_0 =$ 413.80\,Hz and $\dot\nu$\,=\,-0.005\,Hz/s. This signal has a pre-trial p-value of $7.69\times10^{-13}$ (7.16$\sigma$), corresponding to a post-trial p-value of $4.20\times10^{-6}$  (4.6$\sigma$), demonstrating a clear improvement of significance compared to the constant-frequency search.
For comparison, the significance of the signal in this time window for $\dot\nu$\,=\,0 drops to 4.2$\sigma$ (1.85$\sigma$) pre (post) trial. Similarly, the significance of the signal for $\dot\nu$ = -0.005 Hz/s in the time window found in Section \ref{subsec:oscillations} is 5.33$\sigma$ (3.00$\sigma$) pre (post) trial, which is consistent with previous results. This is expected, as a frequency drift becomes more distinguishable from a constant-frequency signal when the analyzed time window is sufficiently long.

\autoref{fig:nu_nudot} (left panel) shows the significance of the signal with respect to its average frequency $\bar\nu$ and its frequency derivative $\dot\nu$, in [$T_0$ + 11\,s, $T_0$ + 62\,s]. 
The central feature, around the point of maximum significance, is well described by a two-dimensional Gaussian function from which we can estimate the uncertainties on both the average frequency and the frequency derivative. We obtain the average frequency in the [$T_0$+11\,s, $T_0$+62\,s] time window $\bar\nu = 413.674 \pm 0.002$\,Hz and a refined value of $\dot\nu = (-4.7 \pm 0.3)\times 10^{-3}$ Hz/s. Uncertainties represent the $1\sigma$ confidence intervals derived from the Gaussian fit. Using these parameters, we fold the burst lightcurve with respect to phase. The resulting folded profile is shown in the right panel of \autoref{fig:nu_nudot}. 

The fractional amplitude of the oscillations can be given by $|A|/\sqrt{2} \bar R $, with $A$ the amplitude of the oscillation obtained by fitting the folded profile to a sinusoidal model, $\bar R$ the average count rate, determined with the method described in \autoref{sec:data}, and a factor $\sqrt{2}$ which allows the fractional amplitude to be given as a rms amplitude. We find a fractional amplitude of $9.6\pm1.3~\%$~rms in the 4-40~keV energy range. In the 4-10~keV and the 10-20~keV energy ranges, the fractional amplitudes are $2.8\pm1.0~\%$ and $27.6\pm6.1~\%$ rms, respectively. In the 20-40 keV range, the flux is too low to allow a reliable computation of the fractional amplitude.
We observe an increase in the fractional amplitude with energy, which could indicate a similar origin of oscillations as those observed in some bursts from 4U~1728-34, as described by \cite{Mahmoodifar_2019}, where oscillations had a strong amplitude in a 6-12~keV energy range but were undetectable in a softer band.
Moreover, the mechanism that generates these high-amplitude oscillations in the tail of the burst is unknown, as it requires a luminosity contrast at the neutron star surface. This issue was described by \cite{Mahmoodifar_2016}, who proposed cooling models that could explain these phenomena.

\begin{figure*}[htbp]
    \centering
    \includegraphics[width=0.49\linewidth]{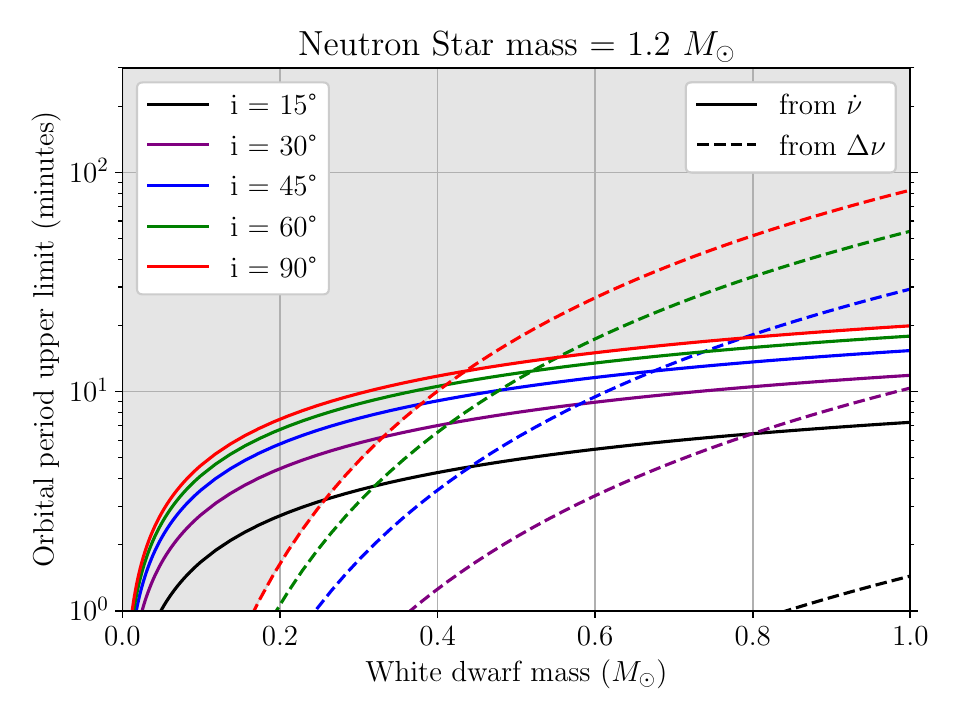}
    \includegraphics[width=0.49\linewidth]{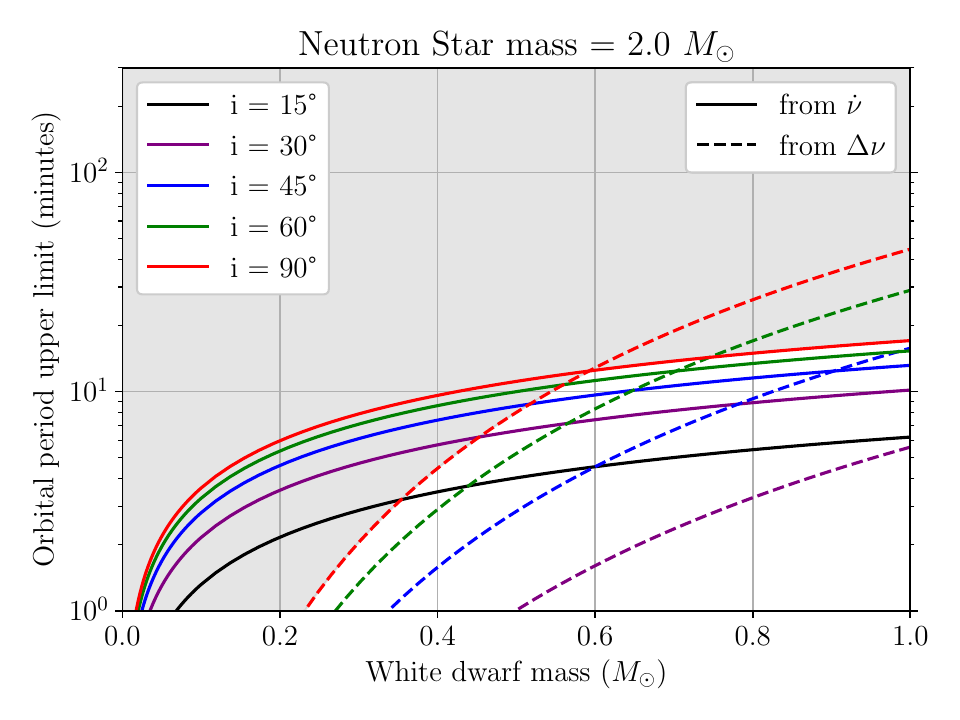}
    \caption{Upper limit of the orbital periods set by the $\Delta\nu$ (dashed lines) and $\dot\nu$ (solid lines) as a function of $M_{\rm{WD}}$, at different inclinations $i$, and for $M_{\rm{NS}} = 1.2 ~ M_\odot$ (left) and $M_{\rm{NS}} = 2.0 ~ M_\odot$ (right). The areas of the parameter space incompatible with the hypothesis that both $\Delta\nu$ and $\dot\nu$ originate from orbital motion are represented in gray.}
    \label{fig:p_orb_max}
\end{figure*}

\section{Discussion}
\label{sec:discussion}

The oscillation frequency that we find with SVOM/ECLAIRs is fully compatible with the signal reported by GECAM \citep{gecam_detection} at 413.63\,Hz, but not the one found with Swift/BAT in a 2006 burst at 414.75\,Hz by \cite{bat_discovery}. To study this discrepancy, we reanalyzed the corresponding archival data using a methodology similar to that described in \autoref{sec:results}. The data were retrieved from the Swift Archive\footnote{\url{https://www.swift.ac.uk/archive/}} under trigger number 234849 (ObsID 00234849000).
We find oscillations $\nu_b = 414.74 \pm 0.02$\,Hz within a time window spanning [25\,s, 32\,s] from the start of the event data, consistent with the value reported in \cite{bat_discovery}. Uncertainty is given as a $1\sigma$, similarly to Section \autoref{sec:drift}. 
No significant frequency variation is observed, and we place a 95\% confidence upper limit on the frequency derivative of $|\dot\nu|\,<\,3.4\times10^{-2}$\,Hz/s. 
The lower precision of the measurement with respect to the one presented in our work mostly comes from the duration of the respective signals: 7 seconds for the initial discovery and 51 seconds for the burst detected by SVOM/ECLAIRs, as a longer signal allows for a more precise measurement of the frequency drift.

The difference between the frequency measured by Swift/BAT and the frequency measured by SVOM/ECLAIRs at the start of the time window is at least $\Delta\nu = 0.9$\,Hz, defined as the distance between their 95\% confidence intervals.
This corresponds to a spin-down rate of $\dot \nu \sim 10^{-9}$\,Hz/s, which is too high to be explained by a spin-down of the neutron star alone. For comparison, in accreting millisecond X-ray pulsars, spin-down rates are typically not greater than $\dot\nu \sim -10^{-13}$\,Hz/s \citep{Patruno_2020}. 

Frequency drifts in burst oscillations are most commonly observed as an increase that stabilizes near the neutron star spin frequency (for example, see Figure~2 of \citealt{2012ARA&A..50..609W}).  However, spin-down episodes have also been observed in 4U~1636--536 \citep{Strohmayer_1999, Miller_2000},  KS~1731--260 \citep{muno_2000}, and  4U~1728--34 \citep{Muno_2002}.
These spin-down episodes exhibit fractional frequency drifts on the order of $\dot\nu/\nu \sim -10^{-3}$\,s$^{-1}$, which is notably larger than the $\dot{\nu}/\nu \sim -10^{-5}$\,s$^{-1}$ reported in this work.  Moreover, the frequency drift here is apparently constant over the entire duration of the oscillations, which is not the case in spin-down episodes previously reported, where the frequency first increased before decreasing. 
The differences in both amplitude and temporal regularity may suggest distinct physical origins for the apparent spin-down reported here compared to those in the earlier cases.

If the frequency drift measured in this work originates from the expansion of a shell at radius $r$, assuming angular momentum conservation, then $\dot{\nu}/\nu~=~2\,\dot{r}/r$. For $r\,=\,6$\,km, this implies an expansion velocity $\dot{r}~\sim~3.4$\,cm/s, which is compatible with a constant radius as seen on \autoref{fig:spectrum_evolution}.
While this scenario is plausible, it fails to explain the steady drift over the entire signal time window; typically, the frequency is expected to increase during the burst tail as the material contracts back onto the neutron star crust \citep{Cumming_2000, Cumming_2002}. 
Therefore, we explore the possibility that the frequency discrepancy between SVOM/ECLAIRs and Swift/BAT $\Delta\nu$ and the frequency drift $\dot\nu$ are caused by a Doppler frequency modulation originating from the orbital motion of the binary. Thus, $\Delta\nu$ would be explained by similar frequencies observed at different orbital phases, and $\dot\nu$ by the evolution of the radial velocity of the neutron star. Under this hypothesis, the comparable frequencies measured by GECAM and ECLAIRs may indicate that both bursts could have happened on a similar orbital phase.

Assuming that either $\Delta\nu$ or $\dot\nu$ is caused by orbital modulation, we aim to constrain the system’s orbital parameters (i.e., the orbital period $P_{\rm{orb}}$, the masses of the neutron star $M_{\rm{NS}}$ and white dwarf $M_{\rm{WD}}$, and the inclination angle $i$). This approach is analogous to the analyses by \cite{Strohmayer_2002} and \cite{2002ApJ...568..279G}, who constrained the masses of the binary components in 4U~1636--53 by inferring the neutron star’s orbital velocity from burst oscillations.

If the orbit is circular, then $\Delta\nu$ provides a lower limit on twice the neutron star's maximum radial velocity along its orbit, $V_{\rm max}$. This, in turn, yields an upper limit on the ratio $P_{\rm orb} / f$:
\begin{equation}    
\begin{split}
    2V_{\rm{max}} = 2\left( \frac{2\pi G f}{P_{\rm{orb}}} \right)^{1/3} > \frac{c\Delta\nu}{\nu}
    \\ \Rightarrow
        \frac{P_{\rm{orb}}}{f} < 2\pi G \left(\frac{c\Delta\nu}{2\nu}\right)^{-3},~
\label{eq:p_orb_deltanu}
\end{split}
\end{equation}
\noindent where $f$ is the system mass function defined as:
\begin{equation}
    f = \frac{\sin^3(i) \cdot M_{\rm{WD}}^3}{\left(M_{\rm{NS}} + M_{\rm{WD}}\right)^2}~~.
\end{equation}

Similarly, the maximum radial acceleration $\dot{V}_{\rm max}$ along the orbit must be at least as large as the one inferred from the observed frequency drift $\dot{\nu}$. This leads to an upper limit on the ratio $P_{\rm orb} / f^{1/4}$:
\begin{equation}
\begin{split}
    \dot V_{\rm{max}} = \frac{2\pi}{P_{\rm{orb}}} V_{\rm{max}} > \frac{c\dot\nu}{\nu} ~~~~~~~~~~~~~
    \\ \Rightarrow
        \frac{P_{\rm{orb}}}{f^{1/4}} < 2\pi G^{1/4}  \left(\frac{c\dot\nu}{\nu}\right)^{-3/4} ~~ .
\label{eq:p_orb_nudot}
\end{split}
\end{equation}

\autoref{fig:p_orb_max} shows the resulting upper limits on the orbital period derived from \autoref{eq:p_orb_deltanu} and \autoref{eq:p_orb_nudot}, plotted as a function of the white dwarf companion mass for different inclination angles, and assuming two limiting neutron star masses: $1.2\,M_\odot$ and $2.0\,M_\odot$.
We show the entire range of inclination, but it is worth mentioning that \cite{Ludlam_2019} finds the inclination of the disk to be between 50\textdegree~and 62\textdegree~from a spectral analysis of NuSTAR data.
Assuming the observed frequency drift is due to orbital motion, the corresponding orbital period would need to be shorter than approximately 20~minutes with ideal system parameters ($M_{\rm{NS}}=1.2\,M_\odot$, $i=90$\textdegree), or shorter than 15.5 minutes with more realistic parameters $M_{\rm{NS}}=2.0\,M_\odot$ \citep{10.1111/j.1365-2966.2009.15430.x} and $i=62$\textdegree~\citep{Ludlam_2019}. These values are in tension with previously reported orbital periods for this source, as seen in \autoref{sec:intro}, but would place the system among the most compact known binaries. 
On the other hand, it follows from \autoref{eq:p_orb_deltanu} that a physically plausible lower limit on $P_{\rm{orb}}$ (arbitrarily taken to 1 minute on \autoref{fig:p_orb_max}) leads to a lower limit on the mass function $f$, i.e., a lower limit on $M_{\rm{WD}}$ for given $M_{\rm{NS}}$ and inclination.
However, these interpretations rely on the assumption that the oscillation frequency is asymptotically stable in the neutron star's frame and remains consistent across bursts. While the first assumption is supported by the behavior observed in several oscillation detections (see, e.g., \cite{Muno_2002}), the second is more uncertain. For example, \cite{Miller_2000} found that frequency variations between bursts of 4U~1636--536 exceeded what would be expected from its orbital motion alone.

\section{Conclusion}
\label{sec:conclusion}

In this Letter, we report the detection by the ECLAIRs coded-mask telescope on board the SVOM satellite of oscillations during a Type-I burst from 4U~0614+091. The oscillations were observed over a 51-second time interval during the burst decay phase, and exhibit a nearly constant frequency drift of \mbox{$\dot\nu = (-4.7 \pm 0.3) \times 10^{-3}$ Hz/s}, with an average frequency of $\bar\nu = 413.674 \pm 0.002$ Hz. The average frequency is close to, but statistically inconsistent with, the oscillation frequency measured during a previous burst observed by Swift/BAT in 2006. We propose that both the observed frequency offset and the steady drift may be explained by Doppler modulation due to orbital motion. 
Under this assumption, the system's orbital period must be shorter than approximately 20 minutes to account for the observed effects. 
Further observations are necessary to refine the system’s orbital parameters and assess the viability of the Doppler-modulation scenario. 
Moreover, an accurate measurement of the short orbital period of this system would make it a validation binary candidate for the Laser
Interferometer Space Antenna \citep[LISA, ][]{kupfer2024lisagalacticbinariesastrometry}.
The measurement of a frequency drift with an unprecedented precision demonstrates the remarkable capabilities of SVOM/ECLAIRs for the study of thermonuclear bursts, thanks to its large FoV and good sensitivity in the 5--15 keV range. 

%% Also note that the akcnowlodgment environment does not support long amounts of text. If you have a lot of people and institutions to acknowledge, do not use this command. Instead, create a new \section{Acknowledgments}.
\begin{acknowledgments}
 This work was supported by CNES, focused on SVOM. The Space-based multi-band Variable Objects Monitor (SVOM) is a joint Chinese-French mission led by the Chinese National Space Administration (CNSA), the French Space Agency (CNES), and the Chinese Academy of Sciences (CAS). We gratefully acknowledge the unwavering support of NSSC, IAMCAS, XIOPM, NAOC, IHEP, CNES, CEA, CNRS, University of Leicester, and MPE. We acknowledge the use of public data from the Swift data archive. 
\end{acknowledgments}


\vspace{5mm}
\facilities{SVOM (ECLAIRs), Swift (BAT)}


\software{numpy \citep{harris2020array},
matplotlib \citep{Hunter_2007}, 
astropy \citep{2022ApJ...935..167A}, 
stingray \citep{matteo_bachetti_2024_13974481, 2019ApJ...881...39H,bachettiStingrayFastModern2024},
Xspec \citep{1996ASPC..101...17A}
          }



\appendix

\section{Validation of the timing analysis method on the Crab pulsar}
\label{app:crab}

At the time of the burst from 4U~0614+091, the Crab Nebula was located in ECLAIRs field of view. The Crab pulsar has a frequency of $\sim 29.5$~Hz, and shows persistent X-ray pulsed emission. This fortuitous observation can be used to validate the analysis procedure described in \autoref{subsec:oscillations}, by comparing its timing results to frequency and frequency derivative retrieved from public data from the Jodrell Bank Crab Pulsar Monthly Ephemeris\footnote{\url{www.jb.man.ac.uk/~pulsar/crab.html}} \citep{Lyne_1993_crab}, issued from radio observations.
In a 15-minute observation overlapping the burst, from \mbox{2025-01-10T15:53} to \mbox{2025-01-10T16:08}, pulses from the Crab were detected at a frequency of 29.5529~Hz. The pulsar frequency, computed from radio ephemerides, was at the time 29.5535~Hz, which gives a relative error of $\sim 10^{-5}$. 
Moreover, no significant frequency drift was observed in this time window, with an upper limit (90\% CL) on the frequency derivative of $4 \times 10^{-6}$~Hz/s, clearly compatible with the actual frequency derivative of the Crab pulsar of $\sim3\times10^{-10}$~Hz/s.
Additional checks were performed on a longer (26 ks) pointing of the Crab taken on 2025-02-06. The measured frequency was compatible with radio ephemeris within a relative error of $\sim 10^{-7}$, and no significant frequency drift was measured, with an upper limit of $3\times 10^{-9}$~Hz/s. 
These results indicate no discrepancy between ECLAIRs timing measurements and radio observations. Thus, no systematic timing bias hindering the result presented in this work was found in the telescope data or in the analysis procedure.


\bibliography{sample631}{}

@software{matteo_bachetti_2024_13974481,
  author       = {Matteo Bachetti and
                  Daniela Huppenkothen and
                  Usman Khan and
                  Himanshu Mishra and
                  Swapnil Sharma and
                  Abbie Stevens and
                  John Swinbank and
                  Amogh Desai and
                  Haroon Rashid and
                  Evandro Martinez Ribeiro and
                  Mihir Tripathi and
                  Gaurav Joshi and
                  Brigitta Sipőcz and
                  Sri Guru Datta Pisupati and
                  Guglielmo Mastroserio and
                  Dhruv Vats and
                  Pranav S and
                  tappina and
                  Saurav Kumar Roy and
                  omargamal8 and
                  Meg Davis and
                  Achilles Rasquinha and
                  Paul Balm and
                  Stuart Mumford and
                  Devansh Shukla and
                  Utkarsh Kumar and
                  parkma99 and
                  Riccardo Campana and
                  Abhinav Kumar and
                  Nitish Garg},
  title        = {StingraySoftware/stingray: Stingray v2.2},
  month        = oct,
  year         = 2024,
  publisher    = {Zenodo},
  version      = {v2.2},
  doi          = {10.5281/zenodo.13974481},
  url          = {https://doi.org/10.5281/zenodo.13974481},
}

@ARTICLE{2019ApJ...881...39H,
          author = {{Huppenkothen}, Daniela and {Bachetti}, Matteo and
                    {Stevens}, Abigail L. and {Migliari}, Simone and {Balm}, Paul and
                    {Hammad}, Omar and {Khan}, Usman Mahmood and {Mishra}, Himanshu and
                    {Rashid}, Haroon and {Sharma}, Swapnil and {Martinez Ribeiro}, Evandro and
                    {Valles Blanco}, Ricardo},
          title = "{Stingray: A Modern Python Library for Spectral Timing}",
          journal = {\apj},
          year = 2019,
          month = aug,
          volume = {881},
          number = {1},
          eid = {39},
          pages = {39},
          doi = {10.3847/1538-4357/ab258d},
          archivePrefix = {arXiv},
          eprint = {1901.07681},
          primaryClass = {astro-ph.IM},
          adsurl = {https://ui.adsabs.harvard.edu/abs/2019ApJ...881...39H}
        }

@article{bachettiStingrayFastModern2024,
         title = {Stingray 2: {{A}} Fast and Modern {{Python}} Library for Spectral Timing},
         shorttitle = {Stingray 2},
         author = {Bachetti, Matteo and Huppenkothen, Daniela and Stevens, Abigail and Swinbank, John and Mastroserio, Guglielmo and Lucchini, Matteo and Lai, Eleonora Veronica and Buchner, Johannes and Desai, Amogh and Joshi, Gaurav and Pisanu, Francesco and Pisupati, Sri Guru Datta and Sharma, Swapnil and Tripathi, Mihir and Vats, Dhruv},
         year = {2024},
         month = oct,
         journal = {Journal of Open Source Software},
         volume = {9},
         number = {102},
         pages = {7389},
         issn = {2475-9066},
         doi = {10.21105/joss.07389},
         urldate = {2024-10-25},
         langid = {english}
        }

@ARTICLE{bat_discovery,
       author = {{Strohmayer}, Tod E. and {Markwardt}, Craig B. and {Kuulkers}, Erik},
        title = "{Discovery of the Spin Frequency of 4U 0614+09 with the Swift Burst Alert Telescope}",
      journal = {\apjl},
         year = 2008,
        month = jan,
       volume = {672},
       number = {1},
        pages = {L37},
          doi = {10.1086/526546},
archivePrefix = {arXiv},
       eprint = {0711.4018},
 primaryClass = {astro-ph},
       adsurl = {https://ui.adsabs.harvard.edu/abs/2008ApJ...672L..37S}
}

@article{gecam_detection,
doi = {10.3847/1538-4357/ac7ff8},
url = {https://dx.doi.org/10.3847/1538-4357/ac7ff8},
year = {2022},
month = {aug},
publisher = {The American Astronomical Society},
volume = {935},
number = {1},
pages = {10},
author = {Chen, Yu-Peng and Li, Jian and Xiong, Shao-Lin and Ji, Long and Zhang, Shu and Peng, Wen-Xi and Qiao, Rui and Li, Xin-Qiao and Wen, Xiang-Yang and Song, Li-Ming and Zheng, Shi-Jie and Song, Xin-Ying and Zhao, Xiao-Yun and Huang, Yue and Lu, Fang-Jun and Zhang, Shuang-Nan and Xiao, Shuo and Cai, Ce and An, Zheng-Hua and Chang, Zhi and Chen, Can and Chen, Gang and Chen, Wei and Dai, Guang-Qi and Du, Yan-Qi and Gao, Min and Gong, Ke and Guo, Dong-Ya and Guo, Zhi-Wei and He, Jian-Jian and Li, Bin and Li, Chao and Li, Chao-Yang and Li, Gang and Li, Jian-Hui and Li, Lu and Li, Qing-Xin and Li, Xiao-Bo and Li, Yan-Guo and Liang, Jing and Liang, Xiao-Hua and Liao, Jin-Yuan and Liu, Jia-Cong and Liu, Xiao-Jing and Liu, Ya-Qing and Luo, Qi and Ma, Xiang and Meng, Bin and Ou, Ge and Shi, Dong-Li and Shi, Jing-Yan and Sun, Gong-Xing and Sun, Xi-Lei and Tuo, You-Li and Wang, Chen-Wei and Wang, Hui and Wang, Huan-Yu and Wang, Jin and Wang, Jin-Zhou and Wang, Ping and Wang, Wen-Shuai and Wang, Yu-Xi and Wen, Xing and Wu, Hong and Xie, Sheng-Lun and Xu, Yan-Bing and Xu, Yu-Peng and Xue, Wang-Chen and Yang, Sheng and Yao, Min and Ye, Jian-Ying and Yi, Qi-Bin and Zhang, Cheng-Mo and Zhang, Chao-Yue and Zhang, Da-Li and Zhang, Fan and Zhang, Fei and Zhang, Hong-Mei and Zhang, Kai and Zhang, Peng and Zhang, Xiao-Lu and Zhang, Yan-Qiu and Zhang, Zhen and Zhao, Guo-Ying and Zhao, Shi-Yi and Zhao, Yi and Zheng, Chao and Zhou, Xing and Zhu, Yue},
title = {GECAM Detection of a Bright Type I X-Ray Burst from 4U 0614+09: Hint for Its Spin Frequency at 413 Hz},
journal = {\apj}
}

@ARTICLE{2022ApJ...935..167A,
       author = {{Astropy Collaboration} and {Price-Whelan}, Adrian M. and {Lim}, Pey Lian and {Earl}, Nicholas and {Starkman}, Nathaniel and {Bradley}, Larry and {Shupe}, David L. and {Patil}, Aarya A. and {Corrales}, Lia and {Brasseur}, C.~E. and {N{\"o}the}, Maximilian and {Donath}, Axel and {Tollerud}, Erik and {Morris}, Brett M. and {Ginsburg}, Adam and {Vaher}, Eero and {Weaver}, Benjamin A. and {Tocknell}, James and {Jamieson}, William and {van Kerkwijk}, Marten H. and {Robitaille}, Thomas P. and {Merry}, Bruce and {Bachetti}, Matteo and {G{\"u}nther}, H. Moritz and {Aldcroft}, Thomas L. and {Alvarado-Montes}, Jaime A. and {Archibald}, Anne M. and {B{\'o}di}, Attila and {Bapat}, Shreyas and {Barentsen}, Geert and {Baz{\'a}n}, Juanjo and {Biswas}, Manish and {Boquien}, M{\'e}d{\'e}ric and {Burke}, D.~J. and {Cara}, Daria and {Cara}, Mihai and {Conroy}, Kyle E. and {Conseil}, Simon and {Craig}, Matthew W. and {Cross}, Robert M. and {Cruz}, Kelle L. and {D'Eugenio}, Francesco and {Dencheva}, Nadia and {Devillepoix}, Hadrien A.~R. and {Dietrich}, J{\"o}rg P. and {Eigenbrot}, Arthur Davis and {Erben}, Thomas and {Ferreira}, Leonardo and {Foreman-Mackey}, Daniel and {Fox}, Ryan and {Freij}, Nabil and {Garg}, Suyog and {Geda}, Robel and {Glattly}, Lauren and {Gondhalekar}, Yash and {Gordon}, Karl D. and {Grant}, David and {Greenfield}, Perry and {Groener}, Austen M. and {Guest}, Steve and {Gurovich}, Sebastian and {Handberg}, Rasmus and {Hart}, Akeem and {Hatfield-Dodds}, Zac and {Homeier}, Derek and {Hosseinzadeh}, Griffin and {Jenness}, Tim and {Jones}, Craig K. and {Joseph}, Prajwel and {Kalmbach}, J. Bryce and {Karamehmetoglu}, Emir and {Ka{\l}uszy{\'n}ski}, Miko{\l}aj and {Kelley}, Michael S.~P. and {Kern}, Nicholas and {Kerzendorf}, Wolfgang E. and {Koch}, Eric W. and {Kulumani}, Shankar and {Lee}, Antony and {Ly}, Chun and {Ma}, Zhiyuan and {MacBride}, Conor and {Maljaars}, Jakob M. and {Muna}, Demitri and {Murphy}, N.~A. and {Norman}, Henrik and {O'Steen}, Richard and {Oman}, Kyle A. and {Pacifici}, Camilla and {Pascual}, Sergio and {Pascual-Granado}, J. and {Patil}, Rohit R. and {Perren}, Gabriel I. and {Pickering}, Timothy E. and {Rastogi}, Tanuj and {Roulston}, Benjamin R. and {Ryan}, Daniel F. and {Rykoff}, Eli S. and {Sabater}, Jose and {Sakurikar}, Parikshit and {Salgado}, Jes{\'u}s and {Sanghi}, Aniket and {Saunders}, Nicholas and {Savchenko}, Volodymyr and {Schwardt}, Ludwig and {Seifert-Eckert}, Michael and {Shih}, Albert Y. and {Jain}, Anany Shrey and {Shukla}, Gyanendra and {Sick}, Jonathan and {Simpson}, Chris and {Singanamalla}, Sudheesh and {Singer}, Leo P. and {Singhal}, Jaladh and {Sinha}, Manodeep and {Sip{\H{o}}cz}, Brigitta M. and {Spitler}, Lee R. and {Stansby}, David and {Streicher}, Ole and {{\v{S}}umak}, Jani and {Swinbank}, John D. and {Taranu}, Dan S. and {Tewary}, Nikita and {Tremblay}, Grant R. and {de Val-Borro}, Miguel and {Van Kooten}, Samuel J. and {Vasovi{\'c}}, Zlatan and {Verma}, Shresth and {de Miranda Cardoso}, Jos{\'e} Vin{\'\i}cius and {Williams}, Peter K.~G. and {Wilson}, Tom J. and {Winkel}, Benjamin and {Wood-Vasey}, W.~M. and {Xue}, Rui and {Yoachim}, Peter and {Zhang}, Chen and {Zonca}, Andrea and {Astropy Project Contributors}},
        title = "{The Astropy Project: Sustaining and Growing a Community-oriented Open-source Project and the Latest Major Release (v5.0) of the Core Package}",
      journal = {\apj},
         year = 2022,
        month = aug,
       volume = {935},
       number = {2},
          eid = {167},
        pages = {167},
          doi = {10.3847/1538-4357/ac7c74},
archivePrefix = {arXiv},
       eprint = {2206.14220},
 primaryClass = {astro-ph.IM},
       adsurl = {https://ui.adsabs.harvard.edu/abs/2022ApJ...935..167A}
}

@article{Buccheri:1983zz,
    author = "Buccheri, R. and others",
    title = "{Search for pulsed gamma-ray emission from radio pulsars in the COS-B data}",
    journal = "\aap",
    volume = "128",
    pages = "245--251",
    year = "1983"
}

@article{Migliari_2010,
doi = {10.1088/0004-637X/710/1/117},
url = {https://dx.doi.org/10.1088/0004-637X/710/1/117},
year = {2010},
month = {jan},
publisher = {The American Astronomical Society},
volume = {710},
number = {1},
pages = {117},
author = {Migliari, S. and Tomsick, J. A. and Miller-Jones, J. C. A. and Heinz, S. and Hynes, R. I. and Fender, R. P. and Gallo, E. and Jonker, P. G. and Maccarone, T. J.},
title = {THE COMPLETE SPECTRUM OF THE NEUTRON STAR X-RAY BINARY 4U 0614+091},
journal = {\apj}
}

@ARTICLE{2010A&A...514A..65K,
       author = {{Kuulkers}, E. and {in't Zand}, J.~J.~M. and {Atteia}, J.~-L. and {Levine}, A.~M. and {Brandt}, S. and {Smith}, D.~A. and {Linares}, M. and {Falanga}, M. and {S{\'a}nchez-Fern{\'a}ndez}, C. and {Markwardt}, C.~B. and {Strohmayer}, T.~E. and {Cumming}, A. and {Suzuki}, M.},
        title = "{What ignites on the neutron star of 4U 0614+091?}",
      journal = {\aap},
         year = 2010,
        month = may,
       volume = {514},
          eid = {A65},
        pages = {A65},
          doi = {10.1051/0004-6361/200913210},
archivePrefix = {arXiv},
       eprint = {0909.3391},
 primaryClass = {astro-ph.HE},
       adsurl = {https://ui.adsabs.harvard.edu/abs/2010A&A...514A..65K}
}

@ARTICLE{2021MNRAS.502.5455A,
       author = {{Arnason}, R.~M. and {Papei}, H. and {Barmby}, P. and {Bahramian}, A. and {Gorski}, M.~D.},
        title = "{Distances to Galactic X-ray binaries with Gaia DR2}",
      journal = {\mnras},
         year = 2021,
        month = apr,
       volume = {502},
       number = {4},
        pages = {5455-5470},
          doi = {10.1093/mnras/stab345},
archivePrefix = {arXiv},
       eprint = {2102.02615},
 primaryClass = {astro-ph.HE},
       adsurl = {https://ui.adsabs.harvard.edu/abs/2021MNRAS.502.5455A}
}

@ARTICLE{2013MNRAS.429.2986M,
       author = {{Madej}, O.~K. and {Jonker}, P.~G. and {Groot}, P.~J. and {van Haaften}, L.~M. and {Nelemans}, G. and {Maccarone}, T.~J.},
        title = "{Time-resolved X-Shooter spectra and RXTE light curves of the ultra-compact X-ray binary candidate 4U 0614+091}",
      journal = {\mnras},
         year = 2013,
        month = mar,
       volume = {429},
       number = {4},
        pages = {2986-2996},
          doi = {10.1093/mnras/sts550},
archivePrefix = {arXiv},
       eprint = {1212.0862},
 primaryClass = {astro-ph.HE},
       adsurl = {https://ui.adsabs.harvard.edu/abs/2013MNRAS.429.2986M}
}

@ARTICLE{2008PASP..120..848S,
       author = {{Shahbaz}, T. and {Watson}, C.~A. and {Zurita}, C. and {Villaver}, E. and {Hernandez-Peralta}, H.},
        title = "{Time-Resolved Optical Photometry of the Ultracompact Binary 4U 0614+091}",
      journal = {\pasp},
         year = 2008,
        month = aug,
       volume = {120},
       number = {870},
        pages = {848},
          doi = {10.1086/590505},
archivePrefix = {arXiv},
       eprint = {0806.1419},
 primaryClass = {astro-ph},
       adsurl = {https://ui.adsabs.harvard.edu/abs/2008PASP..120..848S}
}

@ARTICLE{2014A&A...572A..99B,
       author = {{Baglio}, M.~C. and {Mainetti}, D. and {D'Avanzo}, P. and {Campana}, S. and {Covino}, S. and {Russell}, D.~M. and {Shahbaz}, T.},
        title = "{Polarimetric and spectroscopic optical observations of the ultra-compact X-ray binary 4U 0614+091}",
      journal = {\aap},
         year = 2014,
        month = dec,
       volume = {572},
          eid = {A99},
        pages = {A99},
          doi = {10.1051/0004-6361/201424665},
archivePrefix = {arXiv},
       eprint = {1410.1876},
 primaryClass = {astro-ph.SR},
       adsurl = {https://ui.adsabs.harvard.edu/abs/2014A&A...572A..99B}
}

@article{Nelemans_2004,
   title={Optical spectra of the carbon-oxygen accretion discs in the ultra-compact X-ray binaries 4U 0614+09, 4U 1543−624 and 2S 0918−549},
   volume={348},
   ISSN={1365-2966},
   url={http://dx.doi.org/10.1111/j.1365-2966.2004.07486.x},
   DOI={10.1111/j.1365-2966.2004.07486.x},
   number={1},
   journal={\mnras},
   publisher={Oxford University Press (OUP)},
   author={Nelemans, G. and Jonker, P. G. and Marsh, T. R. and van der Klis, M.},
   year={2004},
   month=feb, pages={L7–L11} }

@ARTICLE{2012ARA&A..50..609W,
       author = {{Watts}, Anna L.},
        title = "{Thermonuclear Burst Oscillations}",
      journal = {\araa},
         year = 2012,
        month = sep,
       volume = {50},
        pages = {609-640},
          doi = {10.1146/annurev-astro-040312-132617},
archivePrefix = {arXiv},
       eprint = {1203.2065},
 primaryClass = {astro-ph.HE},
       adsurl = {https://ui.adsabs.harvard.edu/abs/2012ARA&A..50..609W}
}

@inbook{Patruno_2020,
   title={Accreting Millisecond X-ray Pulsars},
   ISBN={9783662621103},
   ISSN={2214-7985},
   url={http://dx.doi.org/10.1007/978-3-662-62110-3_4},
   DOI={10.1007/978-3-662-62110-3_4},
   booktitle={Timing Neutron Stars: Pulsations, Oscillations and Explosions},
   publisher={Springer Berlin Heidelberg},
   author={Patruno, Alessandro and Watts, Anna L.},
   year={2020},
   month=oct, pages={143–208} }

@ARTICLE{2002ApJ...568..279G,
       author = {{Giles}, A.~B. and {Hill}, K.~M. and {Strohmayer}, T.~E. and {Cummings}, N.},
        title = "{Burst Oscillation Periods from 4U 1636-53: A Constraint on the Binary Doppler Modulation}",
      journal = {\apj},
         year = 2002,
        month = mar,
       volume = {568},
       number = {1},
        pages = {279-288},
          doi = {10.1086/338890},
archivePrefix = {arXiv},
       eprint = {astro-ph/0109294},
 primaryClass = {astro-ph},
       adsurl = {https://ui.adsabs.harvard.edu/abs/2002ApJ...568..279G}
}

@article{Muno_2002,
doi = {10.1086/343793},
url = {https://dx.doi.org/10.1086/343793},
year = {2002},
month = {dec},
publisher = {},
volume = {580},
number = {2},
pages = {1048},
author = {Muno, Michael P. and Chakrabarty, Deepto and Galloway, Duncan K. and Psaltis, Dimitrios},
title = {The Frequency Stability of Millisecond Oscillations in Thermonuclear X-Ray Bursts},
journal = {\apj},
}

@article{10.1093/mnras/sty1404,
    author = {van Doesburgh, Marieke and van der Klis, Michiel and Morsink, Sharon M},
    title = {The highest frequency kHz QPOs in neutron star low-mass X-ray binaries},
    journal = {\mnras},
    volume = {479},
    number = {1},
    pages = {426-434},
    year = {2018},
    month = {05},
    issn = {0035-8711},
    doi = {10.1093/mnras/sty1404},
    url = {https://doi.org/10.1093/mnras/sty1404},
    eprint = {https://academic.oup.com/mnras/article-pdf/479/1/426/25122594/sty1404.pdf},
}

@article{10.1111/j.1365-2966.2009.15430.x,
    author = {Boutelier, Martin and Barret, Didier and Miller, M. Coleman},
    title = {kHz quasi-periodic oscillations in the low-mass X-ray binary 4U 0614+09},
    journal = {\mnras},
    volume = {399},
    number = {4},
    pages = {1901-1906},
    year = {2009},
    month = {11},
    issn = {0035-8711},
    doi = {10.1111/j.1365-2966.2009.15430.x},
    url = {https://doi.org/10.1111/j.1365-2966.2009.15430.x},
    eprint = {https://academic.oup.com/mnras/article-pdf/399/4/1901/17323160/mnras0399-1901.pdf},
}

@article{Ford_1997,
doi = {10.1086/310483},
url = {https://dx.doi.org/10.1086/310483},
year = {1997},
month = {feb},
publisher = {},
volume = {475},
number = {2},
pages = {L123},
author = {Ford, E. and Kaaret, P. and Tavani, M. and Barret, D. and Bloser, P. and Grindlay, J. and Harmon, B. A. and Paciesas, W. S. and Zhang, S. N.},
title = {Evidence from Quasi-periodic Oscillations for a Millisecond Pulsar in the Low-Mass X-Ray Binary 4U 0614+091},
journal = {\apj},
}

@article{Miller_2000,
   title={A Characterization of the Brightness Oscillations during Thermonuclear Bursts from 4U 1636−536},
   volume={531},
   ISSN={1538-4357},
   url={http://dx.doi.org/10.1086/308438},
   DOI={10.1086/308438},
   number={1},
   journal={\apj},
   publisher={American Astronomical Society},
   author={Miller, M. Coleman},
   year={2000},
   month=mar, pages={458–466} }

@ARTICLE{muno_2000,
       author = {{Muno}, Michael P. and {Fox}, Derek W. and {Morgan}, Edward H. and {Bildsten}, Lars},
        title = "{Nearly Coherent Oscillations in Type I X-Ray Bursts from KS 1731-260}",
      journal = {\apj},
         year = 2000,
        month = oct,
       volume = {542},
       number = {2},
        pages = {1016-1033},
          doi = {10.1086/317031},
archivePrefix = {arXiv},
       eprint = {astro-ph/0003229},
 primaryClass = {astro-ph},
       adsurl = {https://ui.adsabs.harvard.edu/abs/2000ApJ...542.1016M}
}

@article{Madej_2014,
   title={X-ray reflection in oxygen-rich accretion discs of ultracompact X-ray binaries},
   volume={442},
   ISSN={0035-8711},
   url={http://dx.doi.org/10.1093/mnras/stu884},
   DOI={10.1093/mnras/stu884},
   number={2},
   journal={\mnras},
   publisher={Oxford University Press (OUP)},
   author={Madej, O. K. and García, J. and Jonker, P. G. and Parker, M. L. and Ross, R. and Fabian, A. C. and Chenevez, J.},
   year={2014},
   month=jun, pages={1157–1165} }

@article{Galloway_2020_minbar,
doi = {10.3847/1538-4365/ab9f2e},
url = {https://dx.doi.org/10.3847/1538-4365/ab9f2e},
year = {2020},
month = {aug},
publisher = {The American Astronomical Society},
volume = {249},
number = {2},
pages = {32},
author = {Galloway, Duncan K. and in ’t Zand, Jean and Chenevez, Jérôme and Wörpel, Hauke and Keek, Laurens and Ootes, Laura and Watts, Anna L. and Gisler, Luis and Sanchez-Fernandez, Celia and Kuulkers, Erik},
title = {The Multi-INstrument Burst ARchive (MINBAR)},
journal = {The Astrophysical Journal Supplement Series},
}

@misc{wei2016,
      title={The Deep and Transient Universe in the SVOM Era: New Challenges and Opportunities - Scientific prospects of the SVOM mission}, 
      author={J. Wei and B. Cordier and S. Antier and P. Antilogus and J. -L. Atteia and A. Bajat and S. Basa and V. Beckmann and M. G. Bernardini and S. Boissier and L. Bouchet and V. Burwitz and A. Claret and Z. -G. Dai and F. Daigne and J. Deng and D. Dornic and H. Feng and T. Foglizzo and H. Gao and N. Gehrels and O. Godet and A. Goldwurm and F. Gonzalez and L. Gosset and D. Götz and C. Gouiffes and F. Grise and A. Gros and J. Guilet and X. Han and M. Huang and Y. -F. Huang and M. Jouret and A. Klotz and O. La Marle and C. Lachaud and E. Le Floch and W. Lee and N. Leroy and L. -X. Li and S. C. Li and Z. Li and E. -W. Liang and H. Lyu and K. Mercier and G. Migliori and R. Mochkovitch and P. O'Brien and J. Osborne and J. Paul and E. Perinati and P. Petitjean and F. Piron and Y. Qiu and A. Rau and J. Rodriguez and S. Schanne and N. Tanvir and E. Vangioni and S. Vergani and F. -Y. Wang and J. Wang and X. -G. Wang and X. -Y. Wang and A. Watson and N. Webb and J. J. Wei and R. Willingale and C. Wu and X. -F. Wu and L. -P. Xin and D. Xu and S. Yu and W. -F. Yu and Y. -W. Yu and B. Zhang and S. -N. Zhang and Y. Zhang and X. L. Zhou},
      year={2016},
      eprint={1610.06892},
      archivePrefix={arXiv},
      primaryClass={astro-ph.IM},
      url={https://arxiv.org/abs/1610.06892}, 
}

@article{Dagoneau_2020,
   title={Onboard catalogue of known X-ray sources for SVOM/ECLAIRs},
   volume={645},
   ISSN={1432-0746},
   url={http://dx.doi.org/10.1051/0004-6361/202038995},
   DOI={10.1051/0004-6361/202038995},
   journal={\aap},
   publisher={EDP Sciences},
   author={Dagoneau, N. and Schanne, S. and Rodriguez, J. and Atteia, J.-L. and Cordier, B.},
   year={2020},
   month=dec, pages={A18} }

@article{Dagoneau_2022,
   title={The SVOM/ECLAIRs image trigger with wavelet-based background correction optimised with a one-year simulation of observations},
   volume={665},
   ISSN={1432-0746},
   url={http://dx.doi.org/10.1051/0004-6361/202141891},
   DOI={10.1051/0004-6361/202141891},
   journal={\aap},
   publisher={EDP Sciences},
   author={Dagoneau, N. and Schanne, S.},
   year={2022},
   month=sep, pages={A40} }

@inproceedings{Godet_2014,
   title={The x-/gamma-ray camera ECLAIRs for the gamma-ray burst mission SVOM},
   volume={9144},
   ISSN={0277-786X},
   url={http://dx.doi.org/10.1117/12.2055507},
   DOI={10.1117/12.2055507},
   booktitle={Space Telescopes and Instrumentation 2014: Ultraviolet to Gamma Ray},
   publisher={SPIE},
   author={Godet, O. and Nasser, G. and Atteia, J.-L. and Cordier, B. and Mandrou, P. and Barret, D. and Triou, H. and Pons, R. and Amoros, C. and Bordon, S. and Gevin, O. and Gonzalez, F. and Götz, D. and Gros, A. and Houret, B. and Lachaud, C. and Lacombe, K. and Marty, W. and Mercier, K. and Rambaud, D. and Ramon, P. and Rouaix, G. and Schanne, S. and Waegebaert, V.},
   editor={Takahashi, Tadayuki and den Herder, Jan-Willem A. and Bautz, Mark},
   year={2014},
   month=jul, pages={914424} }

@Article{Dong2010,
author={Dong, YongWei
and Wu, BoBing
and Li, YanGuo
and Zhang, YongJie
and Zhang, ShuangNan},
title={SVOM gamma ray monitor},
journal={Science China Physics, Mechanics and Astronomy},
year={2010},
month={Jan},
day={01},
volume={53},
number={1},
pages={40-42},
issn={1862-2844},
doi={10.1007/s11433-010-0011-7},
url={https://doi.org/10.1007/s11433-010-0011-7}
}

@inproceedings{gotz_2014,
author = {D. G{\"o}tz and J. Osborne and B. Cordier and J. Paul and P. Evans and A. Beardmore and A. Martindale and R. Willingale and P. O'Brien and S. Basa and C. Rossin and O. Godet and N. Webb and J. Greiner and K. Nandra and N. Meidinger and E. Perinati and A. Santangelo and K. Mercier and F. Gonzalez},
title = {{The microchannel x-ray telescope for the gamma-ray burst mission SVOM}},
volume = {9144},
booktitle = {Space Telescopes and Instrumentation 2014: Ultraviolet to Gamma Ray},
editor = {Tadayuki Takahashi and Jan-Willem A. den Herder and Mark Bautz},
organization = {International Society for Optics and Photonics},
publisher = {SPIE},
pages = {914423},
year = {2014},
doi = {10.1117/12.2054898},
URL = {https://doi.org/10.1117/12.2054898}
}

@INPROCEEDINGS{wu_2012,
       author = {{Wu}, C. and {Qiu}, Y.~L. and {Cai}, H.~B.},
        title = "{SVOM Visible Telescope: Performance and Data Process Scheme}",
    booktitle = {Death of Massive Stars: Supernovae and Gamma-Ray Bursts},
         year = 2012,
       editor = {{Roming}, P. and {Kawai}, N. and {Pian}, E.},
       series = {IAU Symposium},
       volume = {279},
        month = sep,
        pages = {421-422},
          doi = {10.1017/S1743921312013646},
       adsurl = {https://ui.adsabs.harvard.edu/abs/2012IAUS..279..421W}
}

@article{Strohmayer_1999,
doi = {10.1086/312258},
url = {https://dx.doi.org/10.1086/312258},
year = {1999},
month = {aug},
publisher = {},
volume = {523},
number = {1},
pages = {L51},
author = {Strohmayer, Tod E.},
title = {Spin-down of Pulsations in the Cooling Tail of an X-Ray Burst from 4U 1636–53},
journal = {\apj}
}

@article{Cumming_2000,
   title={Rotational Evolution during Type I X‐Ray Bursts},
   volume={544},
   ISSN={1538-4357},
   url={http://dx.doi.org/10.1086/317191},
   DOI={10.1086/317191},
   number={1},
   journal={\apj},
   publisher={American Astronomical Society},
   author={Cumming, Andrew and Bildsten, Lars},
   year={2000},
   month=nov, pages={453–474} }

@article{Li_2018,
doi = {10.3847/1538-4357/aade8e},
url = {https://dx.doi.org/10.3847/1538-4357/aade8e},
year = {2018},
month = {oct},
publisher = {The American Astronomical Society},
volume = {866},
number = {1},
pages = {53},
author = {Li, Zhaosheng and Suleimanov, Valery F. and Poutanen, Juri and Salmi, Tuomo and Falanga, Maurizio and Nättilä, Joonas and Xu, Renxin},
title = {Evidence for the Photoionization Absorption Edge in a Photospheric Radius Expansion X-Ray Burst from GRS 1747–312 in Terzan 6},
journal = {\apj},
}

@INPROCEEDINGS{1996ASPC..101...17A,
       author = {{Arnaud}, K.~A.},
        title = "{XSPEC: The First Ten Years}",
    booktitle = {Astronomical Data Analysis Software and Systems V},
         year = 1996,
       editor = {{Jacoby}, George H. and {Barnes}, Jeannette},
       series = {Astronomical Society of the Pacific Conference Series},
       volume = {101},
        month = jan,
        pages = {17},
       adsurl = {https://ui.adsabs.harvard.edu/abs/1996ASPC..101...17A}
}

@ARTICLE{1996ApJ...469L...9S,
       author = {{Strohmayer}, Tod E. and {Zhang}, William and {Swank}, Jean H. and {Smale}, Alan and {Titarchuk}, Lev and {Day}, Charles and {Lee}, Umin},
        title = "{Millisecond X-Ray Variability from an Accreting Neutron Star System}",
      journal = {\apjl},
         year = 1996,
        month = sep,
       volume = {469},
        pages = {L9},
          doi = {10.1086/310261},
       adsurl = {https://ui.adsabs.harvard.edu/abs/1996ApJ...469L...9S}
}

@article{Cumming_2002,
   title={Hydrostatic Expansion and Spin Changes during Type I X‐Ray Bursts},
   volume={564},
   ISSN={1538-4357},
   url={http://dx.doi.org/10.1086/324157},
   DOI={10.1086/324157},
   number={1},
   journal={\apj},
   publisher={American Astronomical Society},
   author={Cumming, Andrew and Morsink, Sharon M. and Bildsten, Lars and Friedman, John L. and Holz, Daniel E.},
   year={2002},
   month=jan, pages={343–352} }

@article{Strohmayer_2002,
   title={Evidence for a Millisecond Pulsar in 4U 1636−53 during a Superburst},
   volume={577},
   ISSN={1538-4357},
   url={http://dx.doi.org/10.1086/342152},
   DOI={10.1086/342152},
   number={1},
   journal={\apj},
   publisher={American Astronomical Society},
   author={Strohmayer, Tod E. and Markwardt, Craig B.},
   year={2002},
   month=sep, pages={337–345} }

@article{Ludlam_2019,
doi = {10.3847/1538-4357/ab0414},
url = {https://dx.doi.org/10.3847/1538-4357/ab0414},
year = {2019},
month = {mar},
publisher = {The American Astronomical Society},
volume = {873},
number = {1},
pages = {99},
author = {Ludlam, R. M. and Miller, J. M. and Barret, D. and Cackett, E. M. and Coughenour, B. M. and Dauser, T. and Degenaar, N. and García, J. A. and Harrison, F. A. and Paerels, F.},
title = {NuSTAR Observations of the Accreting Atolls GX 3+1, 4U 1702-429, 4U 0614+091, and 4U 1746-371},
journal = {\apj},
}

@inbook{Goldwurm_2022,
   title={Coded Mask Instruments for Gamma-Ray Astronomy},
   ISBN={9789811645440},
   url={http://dx.doi.org/10.1007/978-981-16-4544-0_44-1},
   DOI={10.1007/978-981-16-4544-0_44-1},
   booktitle={Handbook of X-ray and Gamma-ray Astrophysics},
   publisher={Springer Nature Singapore},
   author={Goldwurm, Andrea and Gros, Aleksandra},
   year={2022},
   month=sep, pages={1–57} }

@Article{Hunter_2007,
  Author    = {Hunter, J. D.},
  Title     = {Matplotlib: A 2D graphics environment},
  Journal   = {Computing in Science \& Engineering},
  Volume    = {9},
  Number    = {3},
  Pages     = {90--95},
  publisher = {IEEE COMPUTER SOC},
  doi       = {10.1109/MCSE.2007.55},
  year      = 2007
}

@Article{         harris2020array,
 title         = {Array programming with {NumPy}},
 author        = {Charles R. Harris and K. Jarrod Millman and St{\'{e}}fan J.
                 van der Walt and Ralf Gommers and Pauli Virtanen and David
                 Cournapeau and Eric Wieser and Julian Taylor and Sebastian
                 Berg and Nathaniel J. Smith and Robert Kern and Matti Picus
                 and Stephan Hoyer and Marten H. van Kerkwijk and Matthew
                 Brett and Allan Haldane and Jaime Fern{\'{a}}ndez del
                 R{\'{i}}o and Mark Wiebe and Pearu Peterson and Pierre
                 G{\'{e}}rard-Marchant and Kevin Sheppard and Tyler Reddy and
                 Warren Weckesser and Hameer Abbasi and Christoph Gohlke and
                 Travis E. Oliphant},
 year          = {2020},
 month         = sep,
 journal       = {Nature},
 volume        = {585},
 number        = {7825},
 pages         = {357--362},
 doi           = {10.1038/s41586-020-2649-2},
 publisher     = {Springer Science and Business Media {LLC}},
 url           = {https://doi.org/10.1038/s41586-020-2649-2}
}

@misc{kupfer2024lisagalacticbinariesastrometry,
      title={LISA Galactic binaries with astrometry from Gaia DR3}, 
      author={Thomas Kupfer and Valeriya Korol and Tyson B. Littenberg and Sweta Shah and Etienne Savalle and Paul J. Groot and Thomas R. Marsh and Maude Le Jeune and Gijs Nelemans and Anna F. Pala and Antoine Petiteau and Gavin Ramsay and Danny Steeghs and Stanislav Babak},
      year={2024},
      eprint={2302.12719},
      archivePrefix={arXiv},
      primaryClass={astro-ph.SR},
      url={https://arxiv.org/abs/2302.12719}, 
}

@article{Mahmoodifar_2016,
   title={X-RAY BURST OSCILLATIONS: FROM FLAME SPREADING TO THE COOLING WAKE},
   volume={818},
   ISSN={1538-4357},
   url={http://dx.doi.org/10.3847/0004-637X/818/1/93},
   DOI={10.3847/0004-637x/818/1/93},
   number={1},
   journal={\apj},
   publisher={American Astronomical Society},
   author={Mahmoodifar, Simin and Strohmayer, Tod},
   year={2016},
   month=feb, pages={93} }

@article{Mahmoodifar_2019,
doi = {10.3847/1538-4357/ab20c4},
url = {https://doi.org/10.3847/1538-4357/ab20c4},
year = {2019},
month = {jun},
publisher = {The American Astronomical Society},
volume = {878},
number = {2},
pages = {145},
author = {Mahmoodifar, Simin and Strohmayer, Tod E. and Bult, Peter and Altamirano, Diego and Arzoumanian, Zaven and Chakrabarty, Deepto and Gendreau, Keith C. and Guillot, Sebastien and Homan, Jeroen and Jaisawal, Gaurava K. and Keek, Laurens and Wolff, Michael T.},
title = {NICER Observation of Unusual Burst Oscillations in 4U 1728-34},
journal = {\apj},
}

@article{Lyne_1993_crab,
    author = {Lyne, A. G. and Pritchard, R. S. and Graham Smith, F.},
    title = {23 years of Crab pulsar rotational history},
    journal = {\mnras},
    volume = {265},
    number = {4},
    pages = {1003-1012},
    year = {1993},
    month = {12},
    issn = {0035-8711},
    doi = {10.1093/mnras/265.4.1003},
    url = {https://doi.org/10.1093/mnras/265.4.1003},
    eprint = {https://academic.oup.com/mnras/article-pdf/265/4/1003/3173877/mnras265-1003.pdf},
}
\bibliographystyle{aasjournal}



\end{document}